\documentstyle[aps,prl,epsfig,twocolumn]{revtex}
\begin{document}

\draft

\wideabs{
\title{Phase diagram of $\bbox{S=\frac{1}{2}\ XXZ}$ chain with NNN interaction}
\author{Shunsaku Hirata\cite{e-SH} and Kiyohide Nomura\cite{e-KN}}
\address{Department of Physics, Kyushu University, Fukuoka 812-8581, Japan}
\date{\today}
\maketitle

\begin{abstract}
We study the ground state properties of one-dimensional $XXZ$
model with next-nearest neighbor coupling $\alpha$ and
anisotropy $\Delta$. We find the direct transition between
the ferromagnetic phase and the spontaneously dimerized phase.
This is surprising, because the ferromagnetic phase is
classical, whereas the dimer phase is a purely quantum and
nonmagnetic phase. We also discuss the effect of bond alternation
which arises in realistic systems due to lattice
distortion. Our results mean that the direct transition between the
ferromagnetic and spin-Peierls phase occur.
\end{abstract}

}

\narrowtext

Quantum spin chains have attracted a considerable amount of attention
over the past decades. The main reason for it is the dominant
role played by quantum fluctuation and frustration in these systems \cite{Aff}.
One of the simplest model of them is the spin-$\frac{1}{2}\ XXZ$ chain with
next-nearest-neighbor interaction, described by the Hamiltonian
\begin{eqnarray}
H & = & J\sum_{i}^{L} (h_{i,i+1}+\alpha h_{i,i+2}) , \nonumber\\
h_{i,j} & = & S_i^xS_j^x+S_i^yS_j^y+\Delta S_i^zS_j^z. \label{eq:H}
\end{eqnarray}
In the following, we set NN coupling $J>0$.
This system can also be considered as a coupled chain with
zigzag interchain interaction.
Recently, the physical interest for this model has been renewed,
with relation to experiments;
it is discussed that this system describes
a spin-Peierls compound CuGeO$_3$ \cite{CuGeO3}.

The ground state properties are known for several cases.
When $\alpha =0$, it is integrable by the Bethe ansatz \cite{dCG}.
For $\Delta <-1$, the system becomes the ferromagnetic phase, while
for  $\Delta >1$, it becomes the N\'eel phase.
In the case of $XY$-type anisotropy ($-1\le\Delta\le 1$),
quantum fluctuation destroys the long range order even at the
zero temperature.
The ground state is characterized by
a gapless excitation and algebraic decay of spin correlations
(spin fluid phase).
On the $\alpha =\frac{1}{2}$ line, the other exact results are obtained.
At isotropic point ($\Delta =1$),
ground state is purely dimerized \cite{MG} 
and there exists a finite energy gap above the ground state \cite{AKLT}. 
Recently, it is proven that the ground state is still dimerized
for $\Delta > -\frac{1}{2}$, while for $\Delta <-\frac{1}{2}$
a ferromagnetic phase appears \cite{GMK}.

\begin{figure}
\centering\leavevmode
\epsfig{file=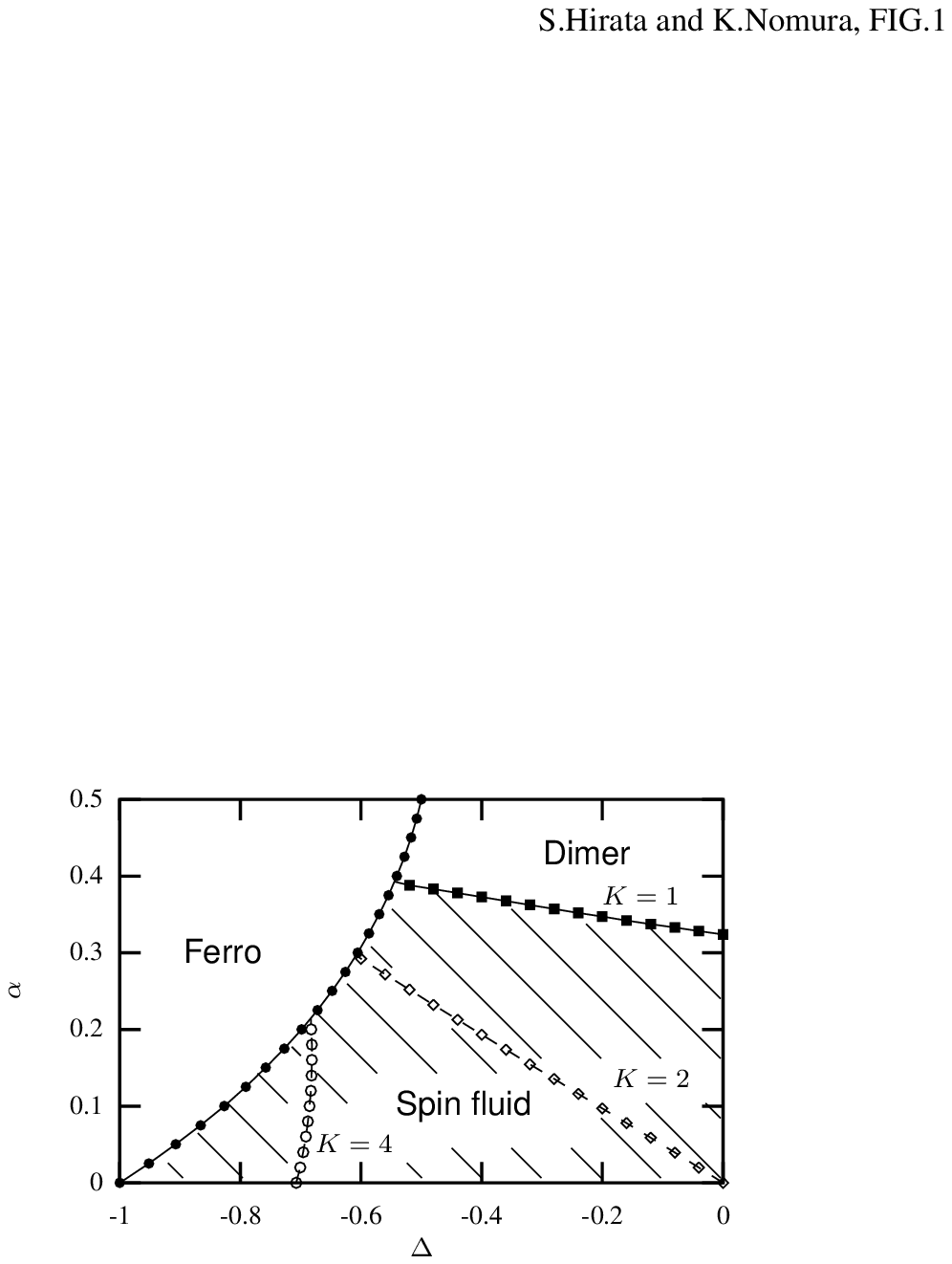,clip}
\caption{Phase diagram of the $S=\frac{1}{2}\ XXZ$ chain with NNN
interaction. Hatched area denotes gapless spin fluid phase.
The dimer-spin fluid boundary is denoted as $K=1$.
Open symbols are related to the instability of spin fluid phase;
$K=2$ ($\diamond$) and $K=4$ ($\circ$).
The lines $K=1,2,4$ terminate at the boundary of ferromagnetic phase,
$(\alpha ,\Delta)=(0.393,-0.545),(0.296,-0.608),(0.214,-0.683)$, respectively.}
\label{FIG1}
\end{figure}

In this letter, we determine the phase diagram of this system in the
region $0\le\alpha\le\frac{1}{2},\ \Delta\le 0$ (see FIG. \ref{FIG1})
by use of diagonalization of finite chain Hamiltonian.
This model has been investigated by many authors \cite{TH,NO}.
The phase diagram of the other region $\Delta\ge0, 0\le\alpha\le\frac{1}{2}$
has been obtained by Nomura and Okamoto \cite{NO}. 
From FIG. \ref{FIG1}, we see that
the transition between the ferromagnetic phase and the dimer phase
occurs in rather broad region in the phase diagram.
This result is surprising, because the ferromagnetic
state is classical, whereas the dimer state has a purely quantum nature.
The magnetic properties may drastically change at the ferro-dimer
transition line.

First, we explain how to determine the boundary of ferromagnetic phase.
The transition from the $S_T^z=\pm L/2$ (Ferro) state to the
$S_T^z=0$ (dimer or spin fluid) state occurs via a level crossing.
Thus, we obtain the boundary of ferromagnetic phase by the level crossing
between the fully magnetized state ($S_T^z=L/2$),
which have the energy eigenvalue $E=\frac{\Delta}{4}(1+\alpha)L$,
and the singlet state ($S_T^z=0$) \cite{NOTE1}.
Then, we compare the numerical result with the variational one.
At $\alpha=\frac{1}{2},\Delta >-\frac{1}{2}$, ground state is purely
dimerized state
\begin{equation}
|\Phi_1\rangle = [1,2]\cdots[L-1,L],\
|\Phi_2\rangle = [2,3]\cdots[L,1],
\end{equation}
where $[i,j]$ is a singlet pair of spin $i$ and $j$.
Therefore, near $\alpha={1\over 2}$, 
we take a trial wave function as
$|\Phi_{var}\rangle=|\Phi_1\rangle+\beta |\Phi_2\rangle $,
and estimate the ground state energy $E_{var}$ variationally.
We find
\begin{equation}
E_{var}/L=-\frac{1}{8}(\Delta +2)
+\frac{B}{1+A}(\Delta+2)\left(\frac{1}{2}-\alpha\right),
\end{equation}
where $A$ and $B$ are positive coefficients of order $2^{-L/2}$.
The ferro-dimer phase boundary is estimated from
$E_{ferro}=E_{var}$, which yields, for $L\to\infty$,
\begin{equation}
\Delta_c=-\frac{2}{2\alpha_c+3}.
\end{equation}
It gives, for $\alpha_c\to\frac{1}{2}$,
\begin{equation}
 \Delta_c=-\frac{1}{2}+\frac{1}{4}\left(\alpha_c-\frac{1}{2}\right)
+O\left(\left(\alpha_c-\frac{1}{2}\right)^2\right).
\end{equation}
It coincides with our data up to the first order of
$\alpha_c-\frac{1}{2}$.

The dimer-spin fluid transition is known to be of
Berezinskii-Kosterlitz-Thouless (BKT) type \cite{BKT}.
It has been difficult to determine the BKT point and critical indices
numerically, since transition occurs with essential
singularity \cite{BM84a}.
In addition, there appears logarithmic correction
as a finite size effect which converges very slowly.
These features make it difficult to determine the transition point accurately.
In fact, a naive application of finite size scaling
methods often lead to a misleading conclusion \cite{BM84a}.
Recently, an efficient method, level spectroscopy, has been proposed
to resolve these difficulties,
based on the renormalization group argument
and symmetry consideration \cite{NO}.

Following the standard steps of Jordan-Wigner transformation and
bosonization, one obtains the 
quantum sine-Gordon Hamiltonian, which describes the large
distance behavior of the Hamiltonian (\ref{eq:H}) \cite{Hal82,KF},
\begin{eqnarray}
H=\frac{1}{2\pi}\int dx\left\{vK(\pi\Pi)^2+\frac{v}{K}(\partial \phi)^2\right\}\nonumber \\
+\frac{v y_{\phi}}{2\pi a^2}\int dx \cos \sqrt{8}\phi ,
\label{eq:sG}
\end{eqnarray}
where $a$ is a short-distance cut off or a lattice spacing,
$v$ is spin wave velocity.
$\Pi $ is momentum density field conjugate to $\phi$,  
$[\phi(x),\Pi(x')]=i\delta(x-x')$. We introduce the field $\theta$
dual to $\phi$, defined as $\partial_x\theta=\pi\Pi$.
We compactified those fields as $\phi\equiv\phi +\frac{\pi}{\sqrt{2}},
\theta\equiv\theta +\frac{\pi}{\sqrt{2}}$.
The spin operators are represented by $\phi,\theta$,
\begin{equation}
S^z_{x/a}\simeq\frac{a}{\sqrt{2}\pi}\partial_x\phi +\frac{1}{\pi}e^{i\pi x/a}\cos\sqrt{2}\phi (x),
\end{equation}
and $S^{\pm}$ is expressed as $e^{\mp i\sqrt{2}\theta}$.
The symmetry operations are interpreted as follow:
translation by one site
\begin{equation}
T_R: \phi\to\phi + \frac{\pi}{\sqrt{2}} , \
\theta\to\theta + \frac{\pi}{\sqrt{2}} ,
\label{eq:T_R}
\end{equation}
spin reversal
\begin{equation}
T: \phi\to -\phi + \frac{\pi}{\sqrt{2}} , \
\theta\to -\theta + \frac{\pi}{\sqrt{2}} ,
\label{eq:T}
\end{equation}
and space inversion is expressed in a similar way,
\begin{equation}
P: \phi\to -\phi+\frac{\pi}{\sqrt{2}}, \
\theta\to\theta + \frac{\pi}{\sqrt{2}} .
\label{eq:P}
\end{equation}

When $y_{\phi}=0$, i.e., the Gaussian, all correlation functions
can be calculated.
The excitation spectrum is gapless (massless) and 
characterized by critical behavior.
When $y_{\phi}\ne 0$, we follow the renormalization group argument.
For $K\simeq 1$ and $g_{\phi}\simeq 0$, we obtain following RG equations 
\cite{sG},
\begin{equation}
\left\{\begin{array}{lcl}
\displaystyle{\frac{dy_0(\ell)}{d\ell}}&=&\displaystyle{-y_{\phi}^2(\ell)}\\
&&\\
\displaystyle{\frac{dy_{\phi}(\ell)}{d\ell}}&=&
\displaystyle{-y_0(\ell)y_{\phi}(\ell)}
\end{array}\right.,
\end{equation}
where $y_0(\ell)$ and $y_{\phi}(\ell)$ are related to bare couplings as
$K=1+\frac{1}{2}y_0(0),\ y_{\phi}(0)=y_{\phi}$,
and the scaling factor $\ell$ is related to the system size $L$
as $\ell =\ln L$.
For $y_0>|y_{\phi}|$, $y_{\phi}$ term is
irrelevant and it is renormalized to zero for $L\to\infty$.
Then, ground state properties are the same as the
case of $y_{\phi}=0$ except $K$ is renormalized,
which correspond to spin fluid phase in the spin system.
While for $y_0<|y_{\phi}|$, $y_{\phi}$ term becomes relevant
and it grows large at large distance limit.
Then, quantum fluctuation is suppressed and spontaneous
symmetry breaking occurs.
The system becomes a N\'eel or dimer phase depending on the sign of
the bare coupling constant $y_{\phi}$ \cite{Hal82}.
The marginal case, $y_0=|y_{\phi}|$, is just critical line and both
couplings are renormalized to zero.
Thus, the critical point is identified as a point at which
the coupling $K$ is renormalized to one.

In the spin fluid region, large distance physics is described by
the Gaussian model with (renormalized) coupling $K$.
Then, the system possesses conformal symmetry \cite{dFMS}.
As a consequence, there exists a state,
in a periodic chain of length $L$,
associated with each primary operator of
scaling dimension $x$ and spin $s$,
whose excitation energy and wave number is given by \cite{Car84},
\begin{equation}
\Delta E=\frac{2\pi v}{L}x,\ k=\frac{2\pi}{L}s.
\label{eq:CFT}
\end{equation}
In our case, the vertex operators \cite{KB} are most significant,
\begin{equation}
O_{n,m}\equiv e^{-i\sqrt{2}n\phi}e^{-i\sqrt{2}m\theta},
\end{equation}
which have,
\begin{equation}
x_{n,m}=\frac{1}{2}\left( Kn^2+\frac{m^2}{K}\right), \
s_{n,m}=nm ,
\label{eq:x_nm}
\end{equation}
where $n,m$ are integers.

From (\ref{eq:CFT}) and (\ref{eq:x_nm}),
we see that, at critical point $K=1$,
some excitations become degenerate, like
$O_{N\acute{e}el}\equiv O_{1,0}+O_{-1,0}$,
$O_{dimer}\equiv O_{1,0}-O_{-1,0}$,
and $O_{doublet}\equiv O_{0,\pm1}$;
these excitations have scaling dimension $x=\frac{1}{2}$.
In the spin system, corresponding excitations are classified as,
$S_T^z=0,\ k=\pi,\ P=-1$ (N\'eel),
$S_T^z=0,\ k=\pi,\ P=1$ (dimer),
$S_T^z=\pm 1,\ k=\pi,\ P=-1$ (doublet)
[cf. (\ref{eq:T_R})--(\ref{eq:P})].
Thus we can determine the transition point from
the crossing of those excitations.

In the sine-Gordon model at criticality,
there appears, as a finite size effect,
logarithmic correction of $O(1/\ln L)$
from the marginally irrelevant field $\cos\sqrt{8}\phi$ \cite{Car86b,AGSZ},
thus the degeneracy of those fields splits.
However, there remains some degeneracy.
The splitting of four $x=\frac{1}{2}$ fields forms
triplet and singlet reflecting implicit $SU(2)$ symmetry
on the BKT transition line \cite{Hal75}.
Therefore, we can eliminate the logarithmic correction 
by choosing member of triplet, as $O_{dimer}$ and $O_{doublet}$,
and we can determine the BKT point accurately.
Besides the logarithmic correction,
there remains a correction of $O(L^{-2})$,
originated from the irrelevant fields ($x=4$) which is not 
included in Hamiltonian (\ref{eq:sG}) \cite{Car84}.

As for the critical indices, logarithmic correction can be eliminated in
the averaged scaling dimension such as 
$(x_{N\acute{e}el}+x_{dimer}+2x_{doublet})/4$,
since the ratio of leading logarithmic correction is known as
$C_{N\acute{e}el}:C_{dimer}:C_{doublet}=3:-1:-1$ reflecting the
implicit $SU(2)$ symmetry of the BKT transition.

\begin{figure}[b]
\centering\leavevmode
\epsfig{file=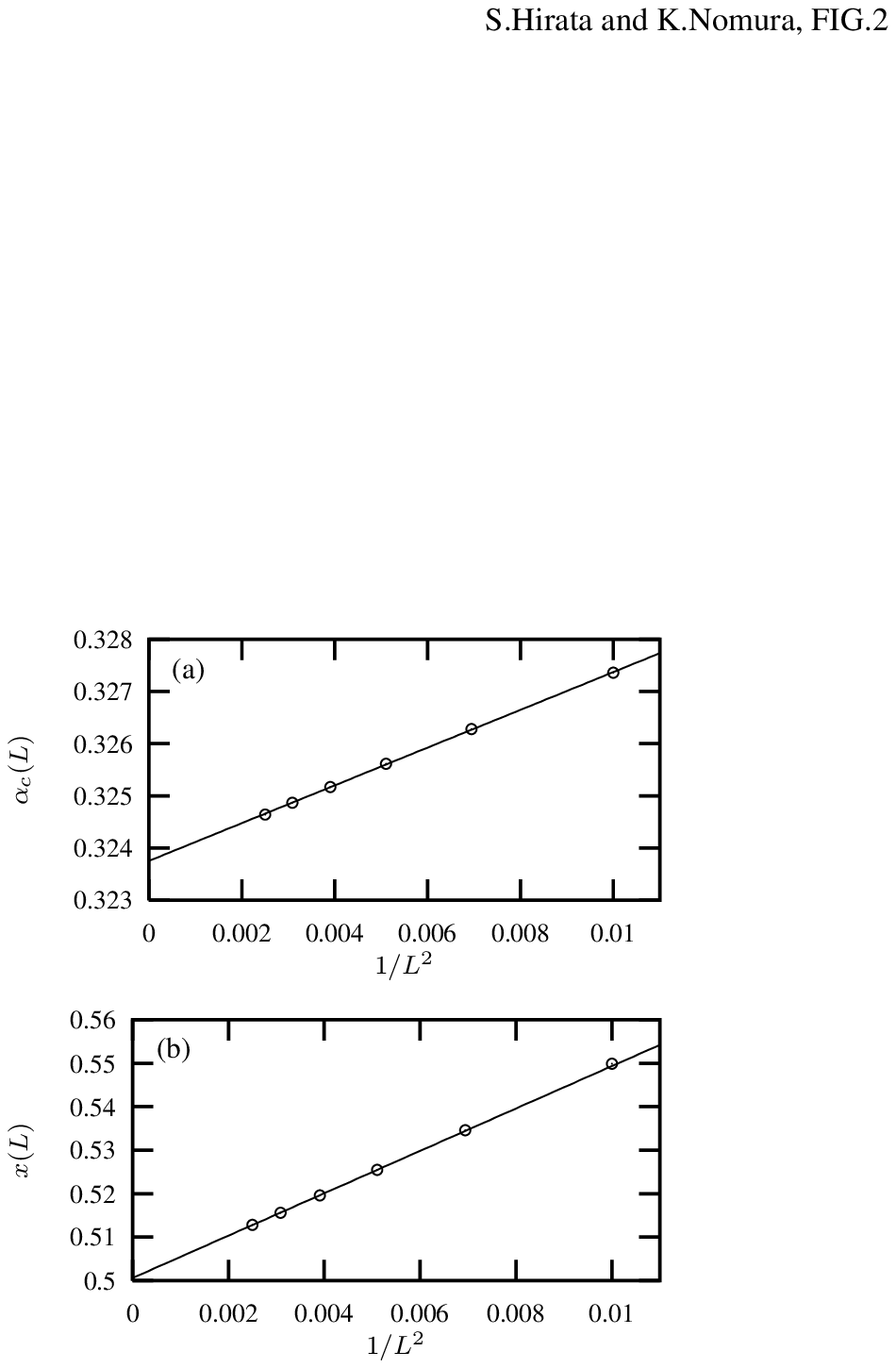,clip}
\caption{(a) Size dependence of crossing point $\alpha_c(\Delta)$ defined as
$\Delta E_{doublet}(\alpha_c,\Delta)=\Delta E_{dimer}(\alpha_c,\Delta)$
and (b) the averaged scaling dimension
$(x_{N\acute{e}el}+x_{dimer}+2x_{doublet})/4$
obtained at BKT point ($K=1$), for $\Delta=0$.}\label{FIG2}
\end{figure}

 Figure \ref{FIG2}(a) shows a size dependence of
crossing points $\alpha_c(L)$.
We see that the leading correction in $\alpha_c(L)$ is $O(L^{-2})$
as expected. We check the averaged scaling dimension
mentioned above, shown in FIG. \ref{FIG2}(b), which converges to the
value $\frac{1}{2}$.
We also estimate the conformal anomaly number $c$ along the BKT line,
from the finite size correction to the ground state energy \cite{BCN},
\begin{equation}
E_0(L)\simeq\epsilon_0 L-\frac{\pi v}{6L}c .
\end{equation}
Our estimate is, $c=0.9956,\ 0.9958, \ 0.9960$ for
$\Delta=0.0,\ -0.2,\ -0.4$, respectively.
We conclude that, together with the averaged scaling dimension,
the system belongs to the same universality class as sine-Gordon model.

In realistic systems, we should consider the lattice distortion or the
staggered field.
The spin fluid phase becomes unstable against such perturbations,
and the system is no longer gapless for $K<4$.
The line $K=2$ is related with the spin-Peierls instability.

For instance, we consider the bond alternation
\begin{equation}
H_{alt.}=\delta\sum_j (-1)^{j}\bbox{S}_j\cdot\bbox{S}_{j+1},
\end{equation}
which has a following bosonized form
\begin{equation}
H_{alt.}=\frac{2g}{(2\pi a)^2}\int dx \sin\sqrt{2}\phi ,
\end{equation}
where $g\sim \delta +O(\delta^3)$, because it must be invariant under
the change $\delta\to -\delta$ 
followed by one site translation $\phi\to\phi+\frac{\pi}{\sqrt{2}}$.
It explicitly breaks the symmetry of one 
site translation,
though the system remains invariant under the two site translation. 
One can see that this form is also invariant under the spin reversal $T$. 
The operator $\sin\sqrt{2}\phi$ has scaling dimension $x=K/2$,
thus it becomes relevant  and causes a gap for $K<4$.

Let us explain how the $K=2$ line is related to 
the spin-Peierls instability.
When $K<4$, $\delta =0$ is just a critical point, and correlation length $\xi$
diverges as
\begin{equation}
\xi\sim \delta^{1/(x-2)}\label{eq:xi},
\end{equation}
which is followed by the RG equation for $\delta$
\begin{equation}
\frac{\partial\delta(\ell)}{\partial\ell}=(2-x)\delta(\ell) +O(\delta^3) .
\end{equation}
To see how the ground state energy changes with the strength of
bond alternation, we consider the following susceptibility
\begin{equation}
\chi\equiv -\frac{\partial^2E_0}{\partial\delta^2},
\end{equation}
where $E_0$ is the ground state energy.
For $\delta\ll 1$, we find
\begin{equation}
\chi/L\sim\xi^{2-2x}\sim\delta^{\frac{2}{2-x}-2}.
\end{equation}
Therefore,
the ground state energy per site $\epsilon_0$ behaves as
\begin{equation}
\epsilon_0(\delta)\approx\epsilon_0(0)-A\cdot\delta^{4/(4-K)} ,
\end{equation}
where $A$ is some positive constant \cite{Cro79}.
The ground state energy decreases as $\delta^a,\ a=4/(4-K)$
due to the formation of singlet pairs.
When $K<2$, it decreases the ground state energy by a sufficient amount
to compensate the energy cost of lattice distortion $\delta^2$,
therefore spin-Peierls instability occurs.

To obtain the $K=2$ and $K=4$ line, we impose the twisted boundary
condition (TBC) with boundary angle $\pi$ which effectively
shifts the value $n$ by an amount $\frac{1}{2}$,
i.e., $n$ changes to $n+\frac{1}{2}$ under 
twisted boundary condition \cite{ABB}.
The excitation $\Delta E_{0,0}^{TBC}$ crosses with 
$\Delta E_{0,1}$ or $\Delta E_{0,2}$ at $K=2$ or $K=4$, respectively.
We also calculate the ratio
\begin{equation}
\Delta E_{1,0}/\Delta E_{0,1}=x_{1,0}/x_{0,1}
\end{equation}
under periodic boundary condition.
According to (\ref{eq:x_nm}), it should be equal to $K^2$.
It can be used to check the result obtained by the method mentioned
above, and we confirmed within $1\%$ accuracy (see TABLE.\ref{table1}).

The $K=1,\ 2,\ 4$ lines terminate at ferro-spin fluid
transition line and they do not merge at the same point.
It indicates that the value $K$ is finite at ferro-spin fluid transition
line except at $\alpha=0$.
This result is somewhat unexpected.
In the spin fluid region, spin correlation behaves as
\begin{eqnarray}
\langle S^z_{x}S^z_{0}\rangle\sim K\left|\frac{x}{a}\right|^{-2}
+const.\times (-1)^{x/a}\left|\frac{x}{a}\right|^{-K}.
\label{eq:zz-corr}
\end{eqnarray}
At first sight, appearance of ferromagnetic order may be
interpreted as $K\to\infty$ where the uniform part of
spin correlation diverges and the staggered one vanishes \cite{Aff}. 
But our result suggests that this picture is incorrect for
$\alpha\ne0$, i.e., $K$ remains finite
at ferro-spin fluid phase boundary except at $\alpha=0,\ \Delta=-1$.

In summary, we determined the phase diagram of one-dimensional $XXZ$ model
with next-nearest neighbor coupling $\alpha$ and anisotropy $\Delta$
in the region $0\le\alpha\le\frac{1}{2},\ \Delta\le0$ (FIG. \ref{FIG1}).
A striking feature of the phase diagram is that
the ferromagnetic phase and the dimer phase share a boundary of
finite length in the phase diagram.
In a realistic situation, where we should include
the lattice distortion in addition to the spin-spin interaction, we will
observe the transition between the ferromagnetic order and the
spin-Peierls order, which can be observed in a broader region 
($K<2$ in FIG. \ref{FIG1}).

The numerical calculation in this work was based on the computer code
TITPACK ver. 2 developed by Professor H. Nishimori.

\begin{table}
\caption{Spin wave velocity $v$, the conformal anomaly number $c$,
and the ratio $x_{1,0}/x_{0,1}$ on $K=2,\ 4$ lines.}
\begin{tabular}{cccc}
&\multicolumn{3}{c}{Extrapolated value}\\
$(\ \Delta ,\ \alpha\ )$& $v$ (Exact\tablenote{Exact value of $v$ is known for $\alpha=0$.})
& $c$ & $x_{1,0}/x_{0,1}$\\ \tableline
$(\ \ 0.00, 0.00\ )$ & $1.000$ ($1$) & $1.000$ & $4.00$\\
$(-0.20 , 0.097)$      & $0.7463$ & $1.000$ & $4.00$\\
$(-0.40 , 0.193)$      & $0.4918$ & $1.000$ & $4.00$\\
$(-0.60 , 0.292)$      & $0.2402$ & $0.996$ & $4.01$\\ \tableline
$(-0.707, 0.00)$       & $0.4731$ ($0.4714$) & $1.000$ & $16.0$\\
$(-0.685, 0.10)$       & $0.3272$ & $0.997$ & $16.0$\\
$(-0.682, 0.20)$       & $0.1990$ & $0.984$ & $15.9$\\
\end{tabular}
\label{table1}
\end{table}



\begin{thebibliography}{10}
\bibitem{NOTE1}
 We have checked whether there are other intermediate phases,
 e.g., partially magnetized phases, and confirmed that there are
 no such a phase.
\end{thebibliography}
\end{document}